\documentclass[prl,aps,twocolumn,groupedaddress,showpacs,floatfix]{revtex4-1}
\usepackage{amsmath,amssymb,multirow,epsfig,bm}
\usepackage{psfrag}
\usepackage{subfigure}

\usepackage[latin1]{inputenc}
\usepackage{amsfonts}
\usepackage{amsmath,amssymb}
\usepackage{graphicx,pst-plot,pst-node}
\usepackage{bm}
\usepackage{subfigure}
\usepackage{hyperref}
\usepackage{bigstrut}
\usepackage{multirow}
\usepackage{notoccite} 
\usepackage{natbib}

\newcommand{\bs}{\boldsymbol}
\newcommand{\sgn}{\text{sgn}}

\newcommand{\tf}{\tilde{\phi}}

\newcommand{\pt}{\partial_{\tau}}
\newcommand{\px}{\partial_x}

\newcommand{\half}{\frac{1}{2}}
\newcommand{\diag}{\text{diag}}

\newcommand{\bsm}{\bs{m}}
\newcommand{\bsn}{\bs{n}}
\newcommand{\bsc}{\bs{c}}
\newcommand{\bsq}{\bs{q}}

\newcommand{\bst}{\bs{t}}

\newcommand{\ttt}{\tilde{t}}

\newcommand{\su}{\mathfrak{su}}
\newcommand{\hsu}{\widehat{\su}}
\newcommand{\ZZ}{\mathbb{Z}}
\newcommand{\tc}{\text{c}}
\newcommand{\tn}{\text{n}}

\begin{document}
\title {Edge Reconstruction in the $\nu=2/3$ Fractional Quantum Hall State}

\author{Jianhui Wang,$^{1,2}$ Yigal Meir,$^{1,3}$ and Yuval Gefen$^2$}
\affiliation{$^1$Department of Physics, Ben-Gurion University of the Negev, Beer Sheva 84105, Israel}
\affiliation{$^2$Department of Condensed Matter Physics, Weizmann Institute of Science, Rehovot 76100, Israel}
\affiliation{$^3$Ilse Katz Institute for Nanoscale Science and Technology, Ben-Gurion University of the Negev, Beer Sheva 84105, Israel}
\begin{abstract}
The edge structure of the $\nu=2/3$ fractional quantum Hall state has been studied for several decades, but recent experiments, exhibiting upstream neutral mode(s), a plateau at a Hall conductance of $\frac{1}{3}( e^2/h)$ through a quantum point contact, and a crossover of the effective charge, from $e/3$ at high temperature to $2e/3$ at low temperature, could not be explained by a single theory. Here we develop such a theory, based on edge reconstruction due to a confining potential with finite slope, that admits  an additional $\nu=1/3$ incompressible strip near the edge. Renormalization group analysis of the effective edge theory due to disorder and interactions explains the experimental observations.
\end{abstract}

\date{\today}

\maketitle

Ever since it has been realized that the low energy dynamics of quantum Hall (and particularly fractional quantum Hall) systems is related to the edge, and that the latter can be described by chiral Luttinger liquids \cite{WenReview1995}, the interest in edge state physics has surged. While the structure of the edge, for simple fractional filling factors, is believed to be well understood, there is recently growing experimental evidence that the situation is much more intriguing and exciting than had been initially believed, when more complex fractions are involved. In this Letter we focus on the $\nu=2/3$ edge. Ostensibly simple, the physics of this edge involves some of the intricacies applicable to other fractions, i.e., edge reconstruction and the emergence of novel elementary excitations (neutral modes in the present case). The major experimental observations addressed by us are as follows: (a) The conductance through a quantum point contact (QPC) exhibits a plateau (versus split-gate voltage) at $G=\frac{1}{3}(e^2/h)$ \cite{Chang1992,Bid2009}. (b) An upstream heating of the QPC has been observed \cite{Bid2010}. (c) The effective charge, detected through shot noise measurements \cite{Bid2009}, crosses over from $e^*=1/3$ at higher temperature to $e^*=2/3$ at lower ones.

Theoretical works have attempted to account for these observations. Observation (a) has been explained \cite{Beenakker1990,Chang1992} by positing that the current at the edge of the $\nu=2/3$ state is carried  by two $\delta \nu=1/3$ edge states, each having the characteristics of a $\nu=1/3$ edge state. While one of these edges is reflected at the QPC, the other is adiabatically transmitted. Observation (b) is associated with a counterpropagating neutral mode, consistent with the Kane-Fisher-Polchinski (KFP) theory \cite{Kane1994}. This theory is based on the renormalization of the original counterpropagating $\delta \nu=1$ and $\delta \nu=-1/3$ edge channels (ECs) \cite{Girvin1984,Macdonald1990} due to interactions and disorder. Observation (c) has been attributed \cite{Ferraro2010,*Ferraro2011} to the competition between relevant operators within the context of the KFP theory. Clearly, there is no single theory currently that can account for all these experimental observations.

\begin{figure}
\begin{center}
\subfigure[]{\label{fig:density-profile}\includegraphics[scale=0.4]{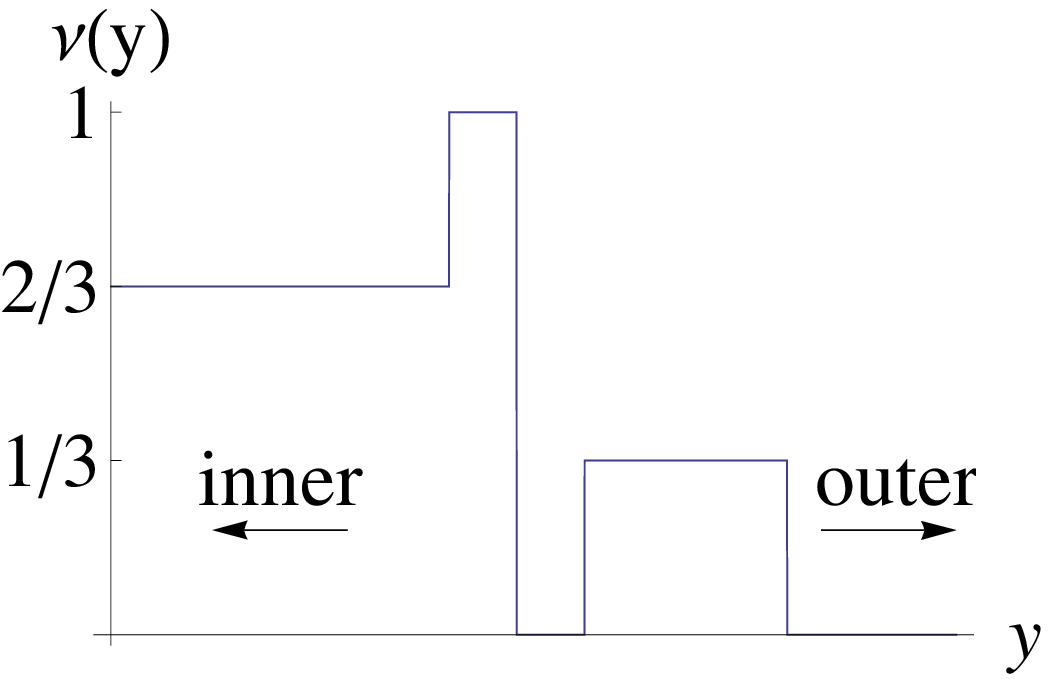}}
\subfigure[]{\label{fig:edge-modes}\includegraphics[scale=0.4]{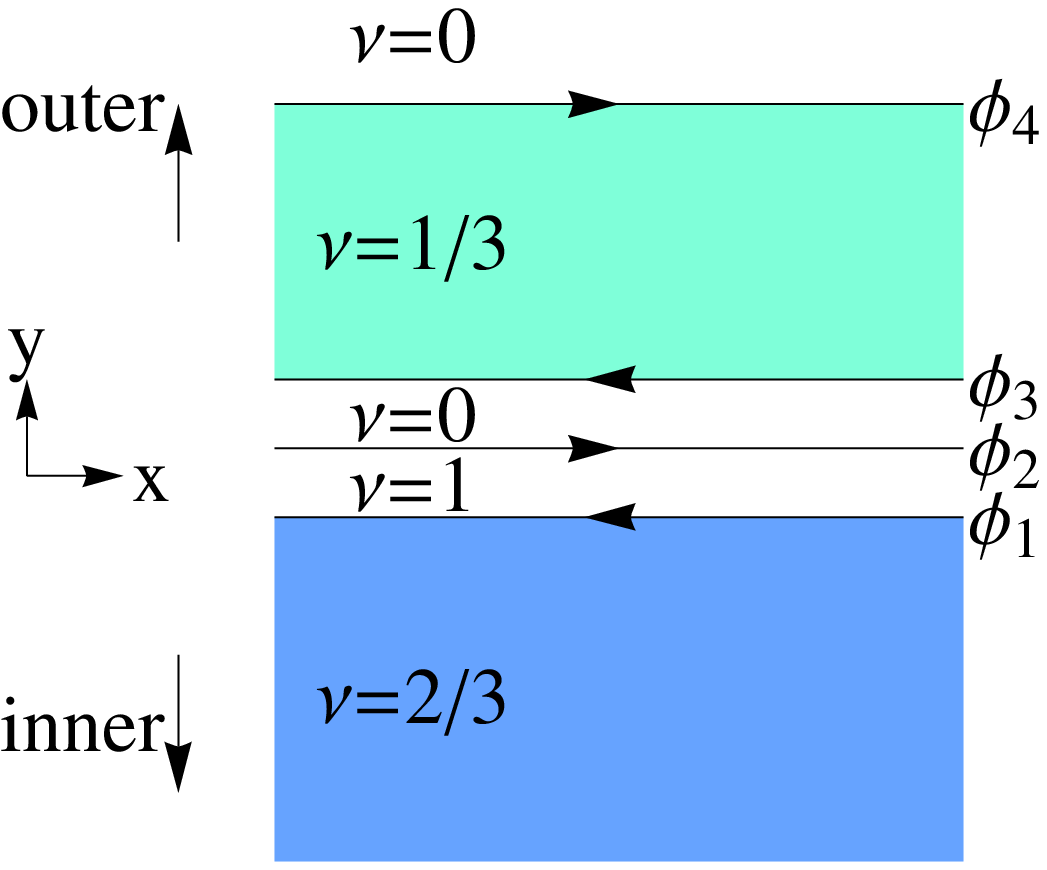}}
\end{center}
\caption{\label{fig:edge3plus1} (color online) The reconstructed edge. (a) The filling factor of the particle-hole conjugate wave function\cite{Girvin1984,Macdonald1990} grows from $\nu=2/3$ in the bulk to $\nu=1$ towards the edge, before falling to zero. In addition, for a smooth potential, the competition between Coulomb and potential energy admits an additional incompressible strip of $\nu=1/3$. (b) The four associated edge modes are depicted, with arrowheads indicating the direction of each mode [downstream (right movers) or upstream (left movers)].
}
\end{figure}

\begin{figure}
\begin{center}
\subfigure[]{\label{fig:bare}\includegraphics[width=4cm]{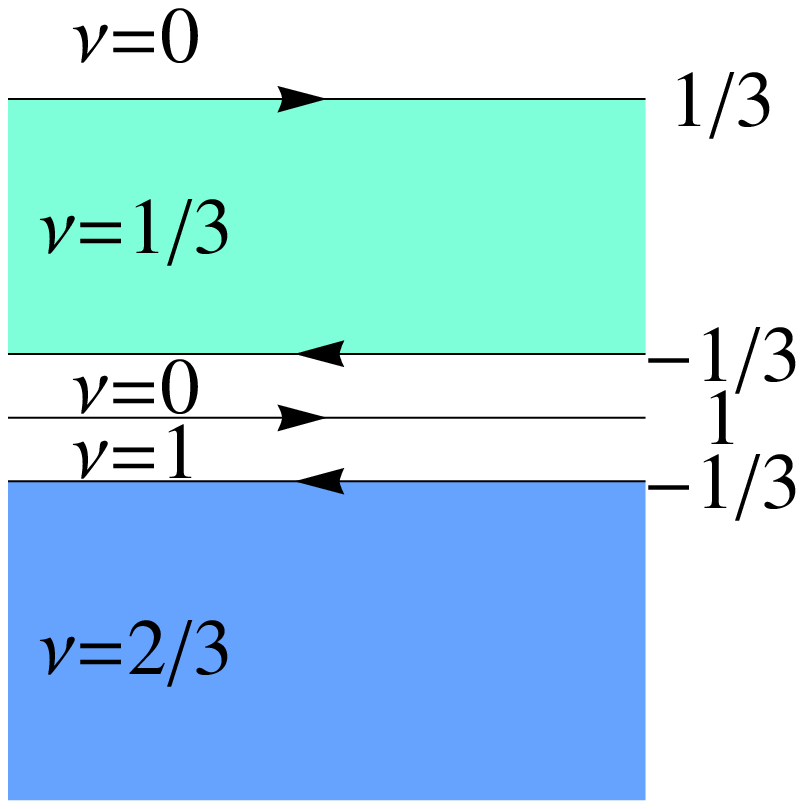}}
\subfigure[]{\label{fig:intrmdt}\includegraphics[width=4cm]{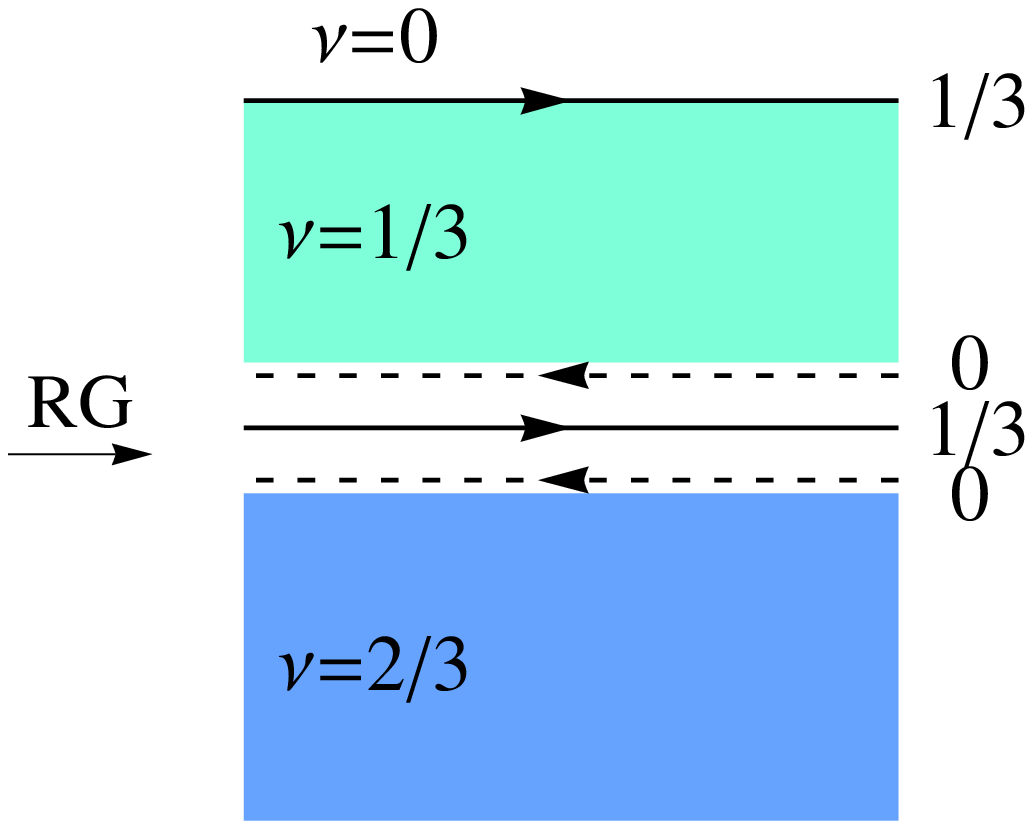}}
\subfigure[]{\label{fig:lowE}\includegraphics[width=4cm]{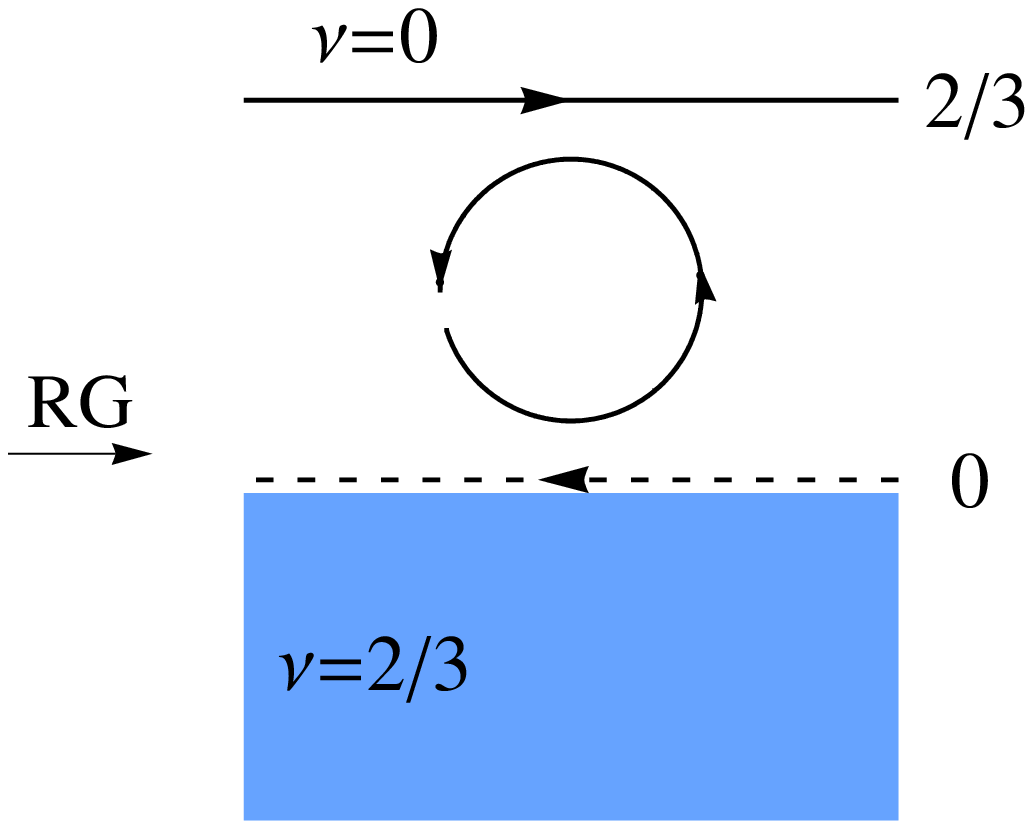}}
\subfigure[]{\label{fig:fsdgrm}\includegraphics[width=4cm]{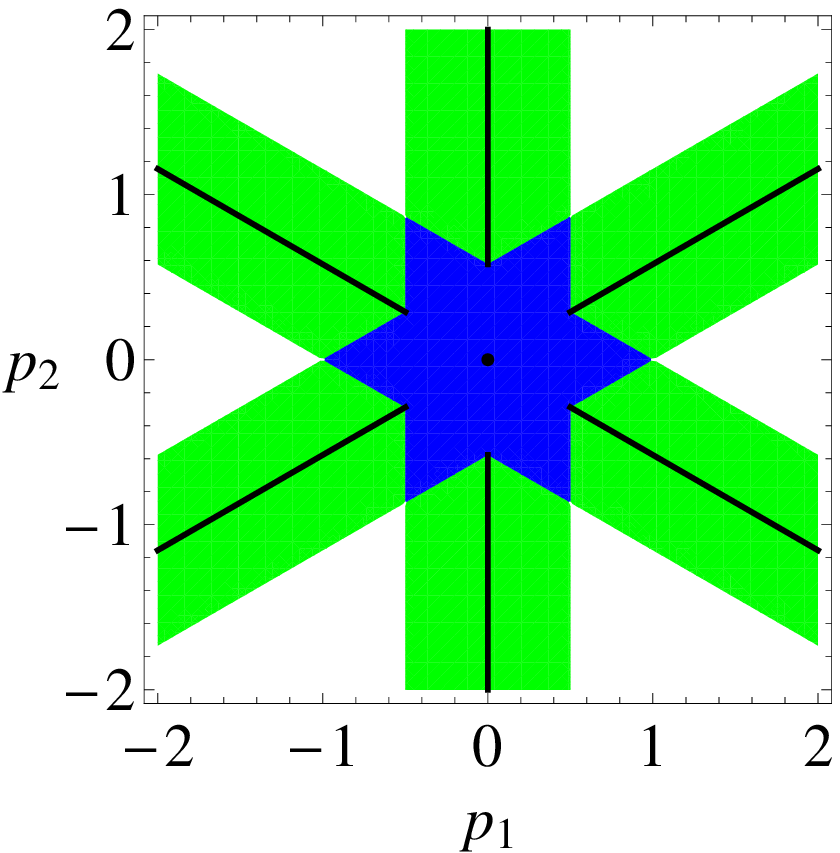}}
\end{center}
\caption{\label{fig:RGRG}(color online) The renormalization group (RG) flow. (a) The bare edge channels. (b) Under RG flow, the three inner channels are renormalized to two upstream neutral modes (as denoted by the dashed lines) and one downstream $1/3$-like charged modes. (The ordering of the three modes is arbitrary.) (c) When temperature is further lowered, the relevant coupling to the outermost mode leads to localization of two modes, as symbolized by the circle, an upstream neutral mode (dashed line) and a downstream $2/3$-like charged mode. (d) Basins of attraction of the various fixed points for the intermediate regime, where the outermost mode is assumed to be decoupled from the other three modes. $p_1$ and $p_2$  parametrize the interaction matrix $V$ \cite{SM}. The green (light gray) strips mark the regions where only one of the three potentially relevant scattering operators [Eq.~(\ref{3rel-op})] is relevant. The blue (dark gray) hexagram is where at least two scattering operators are relevant. The thick black lines in the middle of the green (light gray) strips are fixed lines where there is one upstream neutral mode (in addition to a downstream charged mode and an upstream charged mode) at the interface. The black dot at the origin (the center of the hexagram) is a fixed point where there are two upstream neutral modes [as depicted in (b)].}
\end{figure}

Our theory is based on an edge structure first proposed in Ref.~\cite{Meir1994}. It accounts for all these observations, and provides a general scheme to deal with complex edge structures. According to Ref.~\cite{Meir1994}, the finite slope of the confining potential may lead to a reconstruction of the edge (see also Ref.~\cite{Chamon1994}), resulting in four parallel edge channels. The latter correspond to filling factor discontinuities $\delta \nu=-1/3,+1,-1/3,+1/3$ (moving from the inner edge channel to the outer one, see Fig.~\ref{fig:edge3plus1}). Due to the interplay between the interaction energy and the confining potential, the distance between the inner three modes and the outermost edge could be finite and large, an observation supported by the numerical calculation of Ref.~\cite{Meir1994}. Hence we expect the set of inner ECs to first mix and renormalize among themselves (due to disorder and interaction). In this regime, the emergence of the $1/3$ plateau [observation (a)] is straightforward, as the inner three modes are reflected, while the outer $\delta \nu=+1/3$ mode is transmitted. However, in order to understand the upstream neutral mode [observation (b)] in this regime, one has to carry out a renormalization group (RG) flow calculation within the subspace of the inner three modes. We find (see below) a wide range of parameter space [the green (light gray) strips and the blue (dark gray) hexagram in Fig.~\ref{fig:fsdgrm}] where there are intermediate RG fixed points supporting such an upstream neutral mode (or modes) [see Fig.~\ref{fig:intrmdt}]. At sufficiently low temperatures the relevant coupling to the outer edge begins to play a role, resulting in further renormalizations of the ECs. The emergent picture is that of a $\delta\nu=+2/3$ (downstream) EC, an upstream neutral EC (similar to the KFP theory), and two localized charged modes. The crossover from the intermediate to the low temperature regime explains observation (c).

{\it The model.}-- A general clean edge is described by the action \cite{WenReview1995} \begin{equation}
S_0=\frac{1}{4\pi}\int_{x,\tau}[\px \phi_i K_{ij} i\pt \phi_j+\px \phi_i V_{ij} \px \phi_j]\label{eq-S0}
\end{equation}
(summations over repeated indices are implied). $V$ represents inter- and intrachannel interactions. The bosonic fields $\phi_i$ ($i=1,\dots,N$) are quantized with the usual commutation relations $[\phi_i(x),\phi_j(y)]=i\pi K^{-1}_{ij}\,\sgn(x-y)$.  The symmetric $N\times N$ integer matrix $K$ characterizes the internal structures (the topological orders) in the quantum Hall fluid \cite{WenZee1990,Read1990}. In addition, the charge vector $\bs{t}$ describes the coupling between the charge density in each channel and an external electric field.  The total electric charge density is thus $\rho_{\text{el}}=\frac{1}{2\pi}t_i\partial_x\phi_i$, where the $t_i$'s are the components of $\bs{t}$. The different channels may, in addition, be coupled by impurity scattering \cite{Chamon1994,Kane1994}, with an additional term \begin{equation}
S_{\text{imp}}=\int _{x,\tau}\sum_{\bsn}[\xi_{\bsn}(x)e^{i\bsn\cdot \bs{\phi}(x,\tau)}+\text{H.c.}]\,,
\end{equation}
where $\bs{\phi}$ is the vector whose entries are $\phi_i$, $\bsn$ are constant vectors specifying the scattering processes, $\xi_{\bsn}(x)$ are the random scattering amplitudes. We consider here white noise correlations $\langle\langle \xi_{\bsn}(x)\xi_{\bsn'}(x')\rangle\rangle=D_{\bsn}\delta_{\bsn \bsn'}\delta(x-x')$ \cite{Kane1994,Moore1998,Moore2002}.

The proposed $\nu=2/3$ ground state wave function \cite{Girvin1984,Macdonald1990}, supports two edge modes. This wave function assumes particle-hole symmetry, and is valid for an infinitely sharp confining potential. For a confining potential of a finite slope, it has been demonstrated \cite{Meir1994,Chamon1994} that the competition between the Coulomb energy and the potential energy can lead (and will indeed do so for typical experimental parameters) to the formation of another incompressible strip, of filling factor $\nu=1/3$ in this case. The smoother the potential, the wider the strip, and, as a consequence (see Fig.~\ref{fig:edge3plus1}), the larger the distance between the outer edge mode and the other modes.
Thus, in the present case, $N=4$, and the $K$ matrix may be written \cite{Wen1992} in a diagonal form, with the elements $\{-3,1,-3,3\}$ on the diagonal, and $\bs{t}=(1,\,1,\,1,\,1)^T$.

{\it The intermediate temperature regime.}-- As mentioned above, we  assume that the outermost channel is sufficiently far from the inner three channels, so we can first neglect the interaction and scattering between the former and the latter. The inner three channels are then described by three bosonic fields (cf. Fig.~\ref{fig:edge3plus1}). The $K$ matrix will consist of the upper left $3\times3$  block of the full $K$ matrix, and $\bs{t}=(1,\,1,\,1)^T$.
The allowed scattering operators that are relevant in parts of the parameter space in the RG sense \cite{SM} are $O_{\bsn}=e^{i\bsn\cdot \bs{\phi}}$ with\begin{equation}
\bs{n}_1=\left(\begin{array}{c} 3 \\ 1 \\ 0\end{array}\right)\,,\:\:\bs{n}_2=\left(\begin{array}{c} 0 \\ 1 \\ 3\end{array}\right)\,,\:\:\bsn_3=\left(\begin{array}{c}3\\2\\3\end{array}\right)\,\label{3rel-op}.
\end{equation}

The RG analysis of such a three-channel system has been looked at before \cite{Moore1998,Moore2002}, though not as an intermediate state in a larger space.
Since $K$ is a real symmetric matrix and $V$ a real symmetric positive definite matrix, they can be diagonalized simultaneously \begin{equation}
\Lambda ^TK \Lambda=I_{n^+,n^-}, \, \Lambda ^TV\Lambda=V_{\text{D}}\,\label{diagonalization},\end{equation}
where $V_{\text{D}}$ is a diagonal matrix, and $I_{n^+,n^-}$ is the pseudoidentity matrix with $n^+$ 1's and $n^-$ $-1$'s in the diagonal; for the current problem, $n^+=1$, $n^-=2$. (The structure of the transformation matrix $\Lambda$ is further discussed in Ref.~\cite{SM}.) Equation~(\ref{diagonalization}) amounts to transforming to the basis of the eigenmodes $\bs{\tf}=\Lambda^{-1} \bs{\phi}$ (we use the words ``mode'' and ``channel'' interchangeably). In addition to the amplitudes $D_{\bsn}$, the model has six parameters $V_{ij}\, (i\leq j=1,2,3$), or in the eigenmode basis, $v_i\, (i=1,2,3)$, $\theta$, $p_1$, and $p_2$. The first three are the diagonal elements of  $V_{\text{D}}$ , while the latter three characterize the transformation matrix $\Lambda$  \cite{SM}. In terms of the transformed  fields, the scaling dimensions \cite{SM} can be calculated very easily. The scaling dimension of $e^{i\bsc\cdot \bs{\tf}}$ is simply $\Delta=\half \bsc^2$. The relation between $\bsc$ and $\bsn$ is $\bsc=\Lambda^T \bsn$. Therefore, we find that the scaling dimensions of the three scattering operators specified by Eq.~(\ref{3rel-op}) are \begin{align}
\Delta_{\bsn_1}=&1+2p_1^2\,,\\
\Delta_{\bsn_2}=&\half (2+p_1^2-2\sqrt{3}p_1p_2+3p_2^2)\,,\\
\Delta_{\bsn_3}=&\half (2+p_1^2+2\sqrt{3}p_1p_2+3p_2^2)\,.
\end{align}
Note that  the scaling dimensions only depend on two parameters,  $p_1$ and $p_2$. For an operator to be relevant, its scaling dimension has to be smaller than $3/2$ \cite{Giamarchi1988}. The regions $\Delta_{\bsn_i}<3/2$, are the three green (light gray) stripes in Fig.~\ref{fig:fsdgrm}. The blue (dark gray) hexagram is where at least two of the three operators are relevant.

In regions where none of the three operators is relevant (uncolored regions), the amplitudes of impurity scattering will be renormalized to zero. The other parameters will be somewhat renormalized as well. The result of the RG flow is a clean but nonuniversal interface between the bulk incompressible region and the additional incompressible strip.

In regions of the phase diagram [see Fig.~\ref{fig:fsdgrm}] where one or more impurity operators are relevant, the theory will be renormalized. This means that $\{V_{ij}\}$ will be modified in a nontrivial manner. Alternatively, from the point of view of the eigenmodes, the charge vector $\bs{\ttt}$, and the vectors $\bsc$ in the bosonized form of the electron and quasiparticle tunneling operators will all be renormalized. The coefficients $\bsn$ are, evidently, unchanged.

Within any of the green (light gray) strips, the theory flows to the center line (the thick black lines). On such a fixed line, there is one upstream-moving neutral mode, one upstream-moving charged mode, and one downstream-moving charged mode [not counting the outermost mode, i.e., the topmost one in Fig.~\ref{fig:edge-modes}]. The neutral sector realizes the $\widehat{\mathfrak{su}}(2)_1$ current algebra, so the strong impurity fixed point is exactly solvable, and by simple power counting one can show that it is a stable fixed point.  The charge vector in the eigenmode basis depends on the position on the fixed line. For example, on the $p_1=0$ line, we have (assuming, without loss of generality, $v_2>v_3$) 
\begin{equation}
\bs{\ttt}=(\frac{\sqrt{1+p_2^2}}{\sqrt{3}},\frac{p_2}{\sqrt{3}},0)^T\,.
\end{equation}
Note that the discontinuity in the filling factor (between the bulk value $\frac{2}{3}$ and the value before the outermost channel $\frac{1}{3}$) $\delta \nu=\ttt_1^2-\ttt_2^2-\ttt_3^2=1/3$ is independent of $p_2$ as it should be \cite{Haldane1995,Levkivskyi2009}.

Turning our attention to the hexagram, within this region the RG flows are towards the center (the origin of the $p_1-p_2$ plane). The $(0,0)$ strong impurity fixed point of the three inner modes corresponds to a downstream-moving charged mode (with $\ttt_1=1/\sqrt{3}$), and two upstream neutral modes. The neutral sector realizes the $\widehat{\mathfrak{su}}(3)_1$ current algebra (assuming the neutral mode velocities $v_2$ and $v_3$ are renormalized to the same value in higher order RG, i.e., beyond first order in the impurity strength $D_{\bsn}$'s). We note the similarity between Fig.~\ref{fig:fsdgrm} and the basins of attraction for the principal hierarchy states $\nu=3/5$, $3/11$, etc. \cite{Moore1998}.

It is interesting to compare the emerging picture to Beenakker's model \cite{Beenakker1990}. In our picture, the interface between $\nu=1/3$ and $\nu=2/3$ corresponds to three renormalized modes, $\tilde{\phi}_1$ (downstream $\frac{1}{3}$ charged mode) and two upstream modes. It is $\tf_1$ [assuming we are at the $\hsu(3)_1$ point] that conducts electric current (along with the outermost channel). In Beenakker's picture this $\frac{1}{3}$/$\frac{2}{3}$ interface corresponds to an electric current-conducting compressible strip with no inner structure. When there is a QPC with appropriate split-gate voltage, it is possible to form a region of $\nu=1/3$ under the constriction, with the three inner modes fully reflected [similar to Fig.~2(b) of Ref.~\cite{Beenakker1990}], but with the outermost mode fully transmitted. This would lead to a plateau in two-terminal conductance, $G=\frac{1}{3}(e^2/h)$, in agreement with experiment \cite{Bid2009}. On top of Beenakker's picture of two downstream charged modes, our picture includes the upstream neutral modes [two if the theory flows to the $\hsu(3)_1$ point, one if to the $\hsu(2)_1$ lines], which have also been observed \cite{Bid2010}.

We can also consider the leading backscattering processes at a QPC (assuming the transmission is close to 1). With the assumption that the outermost mode is weakly coupled to the inner three, only the latter participate in the backscattering from the top edge of the sample to the bottom edge. At the $\hsu(3)_1$ point, the scaling dimension of the backscattering processes is simply $2\times \half \bsn^T M_1 M_1^T \bsn$, where $M_1$ is as given in the Supplemental Material \cite{SM}, $\bsn$ is an integer vector specifying the quasiparticle that is backscattered. The factor 2 is due to the fact that now we are considering two edges. The minimum of this expression is 1, corresponding to the most relevant tunneling operators with $\bsn=(0,-1,0)^T$, $(1,2,0)^T$, and $(2,2,3)^T$. All three operators backscatter $e/3$ charge from one edge to the other, so if one measures the shot noise at the weak backscattering limit, the effective charge inferred should be $e/3$.

{\it The low temperature regime.}-- Since the coupling between the outermost channel and the inner channels (through interaction and impurity scattering) is relevant in the RG sense, then even if its bare value is small, which allowed us to neglect it in the intermediate temperature regime above, it becomes significant at lower temperatures. There is a qualitative difference between the ensuing four-channel problem and the three-channel problem considered above. We now discuss the RG flow of the theory, and the ensuing picture approaching the stable low temperature fixed point.  If the impurity amplitudes are small we can still use the same RG equations as for the three-channel problem. We find that under RG, the theory will flow to regions where only one of the (infinitely many) zero-conformal-spin operators is still relevant (actually its scaling dimension flows to zero). This operator involves two of the eigenmodes, say $\tf_1$ and $\tf_4$, whose charge vector components will be the same, $\ttt_1=\ttt_4$. These two modes are unstable against localization. This is consistent with Haldane's null vector criterion about edge stability \cite{Haldane1995}. If the initial point of the RG trajectory has a projection in the $p_1-p_2$ plane close to the origin, the amplitude of yet another operator will also grow and eventually have the form $e^{i\sqrt{2}\tf_3}$. As an impurity operator, it cannot create electric charge. This implies that $\tf_3$ has to be neutral, which is indeed what we see in the result of the numerical integration of the RG equations. In other words, $\tf_3$ is an (upstream-moving) neutral mode, while $\tf_2$ is a (downstream-moving) charged mode ($\ttt_2$ is renormalized to $\sqrt{2/3}$), as in the KFP fixed point \cite{Kane1994}. The neutral sector realizes the $\widehat{\mathfrak{su}}(2)_1$ current algebra. At the low temperature fixed point, the quasiparticles involved in weak backscattering at a QPC are made of the remaining nonlocalized modes only, whose action is \begin{align}
S=&\frac{1}{4\pi}\int_{x,\tau}\px\tf_2(i\pt+v_2\px)\tf_2\label{low-T-action}\\
+&\frac{1}{4\pi}\int_{x,\tau}\px \tf_3(-i\pt+v_3\px)\tf_3+\int_{x,\tau}\left[\xi(x)e^{i\sqrt{2}\tf_3}+\text{H.c}\right]\notag;
\end{align}
i.e., the quasiparticle spectrum should be the same as in the KFP problem (for a more detailed discussion of this point see Ref.~\cite{SM}). It is well known that in that case the minimum scaling dimension for the backscattering processes is $2/3$ corresponding to a process with charge $2e/3$ and two with charge $e/3$. When one lowers the temperature in an experiment, there could be a crossover from the $\hsu(3)_1$ point (where the leading QPC backscattering processes have $e/3$) of the three-channel problem to the KFP point. If at the latter the amplitude of the $2e/3$ process is larger than the $e/3$ processes, one would see a crossover in effective charge from $e/3$ at high temperature to $2e/3$ at low temperature. This has actually been observed \cite{Ferraro2010,*Ferraro2011}.
 
The emergence of two separate fixed points for $\nu=2/3$ is in perfect agreement with the observed power law dependence of the transmission ${\cal T}$ through a QPC in this regime. Figure 4 of Ref.\cite{Bid2009} demonstrates that the transmission starts as a power law at low impinging current (or low voltage) and then saturates at ${\cal T}=1/2$ for high voltage. The low voltage data are well described by the KFP fixed point, which predicts that ${\cal T}$ scales at $V^2$ [see Fig.~1(a) in the Supplemental Material \cite{SM}]. On the other hand, the KFP theory will predict that at high voltage ${\cal T}$ will saturate at unity, with a correction that goes like $V^{-2/3}$, in clear contradiction to the data. Our theory, on the other hand, claims that the high voltage data is described by a different fixed point that supports an outer edge state of $\delta\nu=1/3$. If the three inner edge states are fully reflected, and the backscattering is only due to the outer edge state, the transmission will saturate at ${\cal T}=1/2$, with a correction that behaves like $V^{-4/3}$, in excellent agreement with the data [see Fig.~1(b) in the Supplemental Material \cite{SM}].

The localization transition also has an effect on the current-voltage characteristics (for $eV\gg k_{\text{B}}T$) of tunneling from a Fermi liquid to the edge of the fractional quantum Hall liquid (assuming edge reconstruction also happens in the kind of samples used in that kind of experiment) --
when one lowers the temperature from the range where the effective charge is $e/3$ to where it is $2e/3$, the power in the $I-V$ characteristics should decrease from 3 to 2 \cite{SM,footnote3} (or $3/2$ if the neutral mode is saturated).

{\it Summary.} We have considered a reconstructed edge at $\nu=2/3$, consisting of four edge channels, in order to explain outstanding experimental observations. For a smooth potential,  the interaction and impurity backscattering between the outermost channel and the other channels at the $\frac{1}{3}/\frac{2}{3}$ interface can be neglected at high enough temperatures. Then we have a trivial one-channel problem plus a nontrivial three-channel problem. The latter system may be renormalized by interaction and impurity scattering to high symmetry fixed points. The reflection of the inner three modes by a quantum point contact explains the observed $1/3\ e^2/h$ conductance plateau \cite{Chang1992,Bid2009}, while the emergence of the neutral mode at the symmetric fixed points/lines explains the observed  upstream heat current \cite{Bid2010}. At lower temperature, when the interaction and impurity scattering between the outermost mode and those at the interface cannot be neglected, the system is unstable and will flow towards a stable low temperature fixed point: one pair of counterpropagating modes will localize each other. The two remaining modes form a $\frac{2}{3}$/neutral KFP fixed point \cite{Kane1994}. This results in a crossover in the effective charge, from $e^*=1/3$ at high temperature to a higher effective charge (which could be as high as $e^*=2/3$, depending in the bare amplitudes) at low temperatures, again consistent with experiment \cite{Bid2009}. Our theory is also in excellent agreement  with the the measured scaling of the transmission through a point contact with voltage \cite{Bid2009}. Additionally we made a prediction concerning the crossover in the tunneling exponent into the edge from a Fermi liquid. Recent experiments \cite{Deviatov1,*Deviatov2,*Deviatov3,*Deviatov4,Paradiso2012,Venkatachalam2012} have demonstrated the complexity of the edge even for simple filling factor. Venkatachalam {\it et al}. \cite{Venkatachalam2012}, for example, have observed edge reconstruction at $\nu_{\text{bulk}}=1$, where the filling factor at the edge first goes down to $\nu=2/3$, then to $1/3$. The edge structure from the $\nu=2/3$ region outwards is completely consistent with our picture, giving further credence to the theory presented here.

\begin{acknowledgments}
The authors thank M. Heiblum for discussions.  Support was
provided by the Kreitman Foundation, ISF, BSF, GIF, and MINERVA.
\end{acknowledgments}

\clearpage
\setcounter{equation}{0}
\setcounter{table}{0}
\setcounter{figure}{0}

\begin{widetext}
\begin{section}{Supplemental material}
\begin{subsection}{Relevant Scattering Operators}
We present a brief discussion, based on Ref.~\cite{Moore1998sm,Moore2002sm}, which leads to the identification of the three (possibly) relevant scattering operators. The scattering operators (within a given edge) assume the form $O_{\bsn}=e^{i\bsn\cdot \bs{\phi}}$. The following conditions need to be satisfied: First, from charge conservation, the allowed scattering operators (within a given edge) $O_{\bsn}=e^{i\bsn\cdot \bs{\phi}}$ have to have zero total charge, $\bst^T K^{-1}\bsn=0$. Secondly, the conformal spin $s_{\bsn}= \half \bsn^T K^{-1} \bsn$ has to be an integer, so that the operator $O_{\bsn}$ is bosonic \cite{footnote1}. Finally, for the operator to be relevant, the scaling dimension of $O_{\bsn}$, $\Delta_{\bsn}$, has to be smaller than $3/2$ \cite{Giamarchi1988sm}.

The correlation function (at small space-time separation, compared to the system size), setting all velocities to unity, is given by\begin{equation}
\langle O_{\bsn}(x,\tau)O_{-\bsn}(x',\tau')\rangle \sim \frac{1}{[(\tau+ix)-(\tau'+ix')]^{2h_{\bsn}}[(\tau-ix)-(\tau'-ix')]^{2\bar{h}_{\bsn}}},
\end{equation}
where $h_{\bsn},\,\bar{h}_{\bsn}$ are conformal dimensions. For $\tau=\tau'$, this correlation function is reduced to (up to unimportant prefactors) \begin{equation}
\frac{1}{(x-x')^{2(h_{\bsn}+\bar{h}_{\bsn})}}.\end{equation}
Hence the scaling dimension of $O_{\bsn}$ is $\Delta_{\bsn}=h_{\bsn}+\bar{h}_{\bsn}$. For the concept of conformal spin, we consider the coordinate transformation $\tau+ix\rightarrow z=e^{2\pi (\tau+ix)/L}$, where $L$ is the circumference of the edge, then we have\begin{equation}
\langle O_{\bsn}'(z,\bar{z})O_{-\bsn}'(z',\bar{z}')\rangle=\frac{1}{(z-z')^{2h_{\bsn}}(\bar{z}-\bar{z}')^{2\bar{h}_{\bsn}}},\end{equation}
where the prime on $O_{\bsn}$ denotes that now it is for the coordinate $z$. When $z$ is moved around $z'$, $(z-z')\rightarrow (z-z')e^{i2\pi}$, and the total phase of the RHS of the above correlation function is $e^{-i2\pi 2(h_{\bsn}-\bar{h}_{\bsn})}=:e^{-i2\pi 2s_{\bsn}}$, i.e. $s_{\bsn}\equiv h_{\bsn}-\bar{h}_{\bsn}$ \cite{Tsvelik}. 
From the definition of $s_{\bsn}$ and $\Delta_{\bsn}$, it is obvious that $|s_{\bsn}|\leqslant \Delta_{\bsn}$, with the equality satisfied only when $O_{\bsn}$ is chiral or anti-chiral. Hence $s_{\bsn}$ is further restricted to 0 and $\pm 1$ for relevant scattering operators. For the basis used here (the so-called symmetric basis), the components of $\bsn$ are integers (the quasiparticle spectrum is determined by the locality condition, see e.g. \cite{Levkivskyi2009sm}), then the above conditions give three solutions\begin{equation}
\bs{n}_1=\left(\begin{array}{c} 3 \\ 1 \\ 0\end{array}\right)\,,\:\:\bs{n}_2=\left(\begin{array}{c} 0 \\ 1 \\ 3\end{array}\right)\,,\:\:\bsn_3=\left(\begin{array}{c}3\\2\\3\end{array}\right)\,\label{3rel-op-sm}.
\end{equation}
The conformal spin is $-1$ for all three operators.
\end{subsection}

\begin{subsection}{RG parameters}
The problem at hand contains a large number of parameters. Here we show that the interesting features of the phase diagram may be described in terms of only two parameters, $p_1$ and $p_2$, and how the latter are related to the original parameters. We also point out certain properties of the matrix $\Lambda$, which transforms the original bosonic fields into the eigenmodes.

For each impurity scattering process, there is one parameter that describes the strength of the impurity, $D_{\bsn}$. Apart from these, there are $d(d+1)/2$ interaction parameters, where $d=\dim K$ is the number of edge channels. For these one may choose the elements of the real symmetric matrix $V$, $v_{ij}$ with $i,j=1,\dots,d$, $i\leq j$, which describe the interactions between the channels and within each channel. Or, one may choose the $d$ eigenvalues of $V$, $v_i$, $i=1,\dots,d$, plus the parameters of $\Lambda$. It is convenient to decompose $\Lambda$ into three factors, $\Lambda=PM_1M_2$. $P$ permutate the original modes so that all the right-moving modes appear before the left-moving ones [e.g. in the intermediate regime $P$ permutes $\phi_1$ and $\phi_2$ so that $K$ is transformed to $K'=\diag(1,-3,-3)$]. $M_1$ diagonalizes $K'$, $M_1^TK'M_1=I_{n^+,n^-}$. $P$ and $M_1$ do not include RG parameters. $M_2$ diagonalizes $M_1^TVM_1$ without affecting the diagonalization of $K$, $M_2^TI_{n^+,n^-}M_2=I_{n^+,n^-}$, i.e. $M_2\in \text{SO}(n^+,n^-)$, the proper pseudo-orthogonal group \cite{Kane1995,Moore1998sm,Moore2002sm}. Furthermore, $M_2=BR$ \cite{Moore1998sm,Moore2002sm}, where $B$ and $R$ contain the noncompact and compact generators respectively (Cartan's theorem). More precisely, there are $n_+ n_-$ parameters for the $B$ matrix (say $p_r$, $r=1,\dots,n_+n_-$)
, and $n_+(n_+-1)/2+n_-(n_--1)/2$ parameters for the $R$ matrix [say $\theta_s$, $s=1,\dots,n_+(n_+-1)/2+n_-(n_--1)/2$]. With the constraint $n_++n_-=d$, one can see that the numbers of the $v_i$'s, the $p_r$'s and the $\theta_s$'s add up to $d(d+1)/2$ as they should be. For the purpose of the RG calculation, the latter parametrization is much more convenient. To lowest order in impurity strengths $D_{\bsn}$, the eigenmode velocities $v_i$ are not renormalized. All the other parameters are renormalized. The scaling dimensions of the impurities operators  depend only on the $p$'s. One can express $v_{ij}$ as functions of $v_i$, $p_r$, and $\theta_s$, since \begin{equation}
V=\Lambda^{-1\,T}(p_r,\theta_s)V_D(v_i)\Lambda^{-1}(p_r,\theta_s)\label{V-breakdown},\end{equation}
\begin{equation}
\Lambda(p_r,\theta_s)=PM_1B(p_r)R(\theta_s)\label{Lambda-parametrization}.\end{equation}
For the intermediate regime, $d=3$, $n_+=1$, $n_-=2$, and the explicit forms of the matrices on the RHS of Eq.~(\ref{Lambda-parametrization}) are
\begin{align}
&M_1=\left(\begin{array}{ccc}\sqrt{3}&-\frac{1}{\sqrt{2}} & -\sqrt{\frac{3}{2}} \\ -\frac{1}{\sqrt{3}} & \frac{1}{\sqrt{2}} & \frac{1}{\sqrt{6}} \\ -\frac{1}{\sqrt{3}} & 0 & \sqrt{\frac{2}{3}}\end{array}\right),\label{expr-M1}  R=\left(\begin{array}{ccc}1 & 0 & 0 \\ 0 & \cos\theta & -\sin\theta \\ 0 & \sin\theta & \cos\theta \end{array}\right),\\
&B=B^T=\left(\begin{array}{ccc}\gamma & p_1 & p_2 \\ & 1+p_1^2(\gamma-1)/p^2 & p_1p_2(\gamma-1)/p^2 \\ & & 1+p_2^2(\gamma-1)/p^2\end{array}\right)\,\label{B-matrix},\\
&p^2=p_1^2+p_2^2\,,\:\gamma=\sqrt{1+p^2}\,\label{p2-gamma}.
\end{align}
These forms of $M_1$ and $B$ are chosen to simplify  the expressions of the scaling dimensions and the diagram for the basins of attraction. For the actual RG calculation, another form of $B$ may be more convenient, namely \[
B=\left(\begin{array}{ccc}0 & b_{12} & b_{13} \\ b_{12} & 0 & 0 \\ b_{13} & 0 & 0\end{array}\right).\]
Below we stick to the version with the $p$'s.

In principle one can invert Eq.~(\ref{V-breakdown}) to express $v_i$, $p_r$, and $\theta_s$ in terms of $v_{ij}$. However, we do not see any benefit for such an exercise. As we already mentioned, $v_i$, $p_r$, and $\theta_s$ are much more suitable than $v_{ij}$ for the RG calculation. For the experimental point of view, the $v_{ij}$ cannot be measured any way, let alone tuned. From the theoretical point of view, there is no reason to think $v_{ij}$ are more fundamental than $v_i$, $p_r$, and $\theta_s$. Finally, we want to mention that in general the expression for say $p_r(v_{ij})$ would be very complicated, if one can derive them at all. However, it is easy to determine the values of $p_r$'s if one is given numerical values for the $v_{ij}$'s.

From Eq.~(\ref{Lambda-parametrization}), we see that $B=M_1^{-1}P\Lambda R^{-1}$. Transposing this equation and multiplying it with the transpose, noticing $R$ is orthogonal and $B$ is symmetric, we find $B^2=M_1^{-1}P\Lambda \Lambda^TPM_1^{-1T}$. As soon as we determine $\Lambda$ numerically \cite{footnote2}, $p_r$ can be read off, since the LHS is (in the intermediate regime)\[
B^2=\left(\begin{array}{ccc}1+2p^2 & 2p_1\gamma & 2p_2\gamma \\ & 1+2p_1^2 & 2p_1p_2 \\ & & 1+2p_2^2\end{array}\right)\,.\]

\end{subsection}

\begin{subsection}{Low temperature fixed point}
In this section we give a more detailed discussion of the low temperature fixed point. In particular, we discuss the quasiparticle spectrum and determine the $I-V$ characteristics of tunneling from a Fermi liquid to the edge.

As implied in the main text, after the theory flows away from the intermediate fixed point under RG, the action acquires the form
\begin{align}
S=&\frac{1}{4\pi}\int_{x,\tau}\px\tf_2(i\pt+v_2\px)\tf_2\label{low-T-action-sm}\\
+&\frac{1}{4\pi}\int_{x,\tau}\px \tf_3(-i\pt+v_3\px)\tf_3+\int_{x,\tau}\left[\xi(x)e^{i\sqrt{2}\tf_3}+\text{h.c}\right]\notag \\
+&\frac{1}{4\pi}\int_{x,\tau}[ \px \tf_1(i\pt+v_1\px)\tf_1+\tf_4(-i\pt+v_4\px)\tf_4]+\int_{x,\tau}\left[\xi'(x)e^{ic(\tf_1+\tf_4)}+\text{h.c.}\right]\notag
\end{align}
where the last line describes the two modes that will localize each other, the first two lines are the remaining unlocalized modes at the low temperature fixed point. More specifically, the first line is the charged mode and the second line the neutral sector. Since the transport properties are determined by the unlocalized modes, we will forget about the localized modes and rename the charged mode ($\tf_2$) and the neutral mode ($\tf_3$) $\tf_{\tc}$ and $\tf_{\tn}$ respectively.

We know that the charge vector is $\bs{\ttt}=(\sqrt{\frac{2}{3}},0)^T$. We also know that they are downstream-moving and upstream-moving respectively. In other words, the diagonalized $K$-matrix is \begin{equation}
\sigma\equiv \diag(\sigma_{\tc},\sigma_{\tn})=\diag(1,-1).\end{equation}
However, in general, this knowledge does not completely specify the edge theory (let us forget the bulk topological order for the moment). One can construct different edge theories by choosing different electron operators and the corresponding quasiparticle spectrum \cite{Levkivskyi2009sm}. The first step is to find two independent electron vectors $\bsq_{\alpha}$ (the electron operators are given by $e^{i\bsq_{\alpha}\cdot \bs{\tf}}$; here Greek indices, running over 1, 2, denote the different electron operators while Latin indices, running over c, n, denote components of vectors) to span the electron lattice (i.e. the lattice $\ZZ \bsq_1+\ZZ \bsq_2$). The first requirement on $\bsq_{\alpha}$ is that $\theta_{\alpha}=\pi \sum_i\sigma_iq_{\alpha i}q_{\alpha i}$ be an odd integer times $\pi$ (so that the electron operators are fermionic), i.e. $q_{\alpha \tc}^2-q_{\alpha \tn}^2=n_{\text{odd},\alpha}$. The second requirement on $\bsq_{\alpha}$ is that the corresponding charge should be one, \begin{equation}
\sum_i\sigma_i\ttt_iq_{\alpha i}=1\label{unit-charge-condition}.\end{equation}
This means $q_{\alpha \tc}=\sqrt{3/2}$, then $q_{\alpha \tn}=\pm \sqrt{\frac{3}{2}-n_{\text{odd},\alpha}}$. The third requirement on the $\bsq_{\alpha}$'s is that $q_{1\tc}q_{2\tc}-q_{1\tn}q_{2\tn}=\text{integer (odd or even)}$. Then if we limit ourselves to $n_{\text{odd},\alpha}\geq -7$ ($\pi n_{\text{odd},\alpha}$ is the statistical phase associated with exchanging two $\bsq_{\alpha}$ electrons, and we stipulate that the magnitude of this phase should not be too large), there are at most 8 inequivalent lattices. The third of these has the same $K$ matrix as the second one (for the calculation of the $K$-matrix see below), but the lattices are not exactly the same, they are mirror images about the horizontal axis; one can be changed to the other by $\tf_{\tn}\rightarrow -\tf_{\tn}$, which does not change the charge of the vertex operator. Hence we will identify Model 3 with Model 2. Model 4 has the same $K$ as Model 1, and the lattices are the same, so they are really the same model. So we have six different models with $n_{\text{odd}}\geq -7$: \begin{align*}
\text{Model 1: }&\bsq_1=\left(\sqrt{\frac{3}{2}},\frac{1}{\sqrt{2}}\right),\,\bsq_2=\left(\sqrt{\frac{3}{2}},\frac{3}{\sqrt{2}}\right)\\
\text{Model 2: }&\bsq_1=\left(\sqrt{\frac{3}{2}},\frac{1}{\sqrt{2}}\right),\,\bsq_2=\left(\sqrt{\frac{3}{2}},-\frac{3}{\sqrt{2}}\right)\\
\text{Model 5: }&\bsq=\left(\sqrt{\frac{3}{2}},\pm\sqrt{\frac{5}{2}}\right)\\
\text{Model 6: }&\bsq=\left(\sqrt{\frac{3}{2}},\pm\frac{3}{\sqrt{2}}\right)\\
\text{Model 7: }&\bsq=\left(\sqrt{\frac{3}{2}},\pm\sqrt{\frac{13}{2}}\right)\\
\text{Model 8: }&\bsq=\left(\sqrt{\frac{3}{2}},\pm\sqrt{\frac{17}{2}}\right)\end{align*}
However, we know that $e^{i\sqrt{2}\tf_{\tn}}$ is an allowed quasiparticle operator, which means that it should be local with respect to the electron operators. This in turn means that \begin{equation}
\bsq_{\alpha}\,\sigma\left(\begin{array}{c}0 \\ \sqrt{2}\end{array}\right)\in \ZZ,\; \alpha=1,\,2\label{locality-condition}.\end{equation}
Hence only Models 1, 2 and 6 are possible. More generally, all quasiparticles $e^{i\bs{p}\cdot \bs{\tf}}$ should satisfy Eq.~(\ref{locality-condition}) with $(0,\sqrt{2})^T$ replaced by $\bs{p}$. The solution of this equation is  $p_j=\sigma_j\sum_{\beta}q^{-1}_{j\beta}n_{\beta}=\sum_{\beta}n_{\beta}\sigma_jq^{-1}_{j\beta}$, where $q^{-1}$ is the inverse of the matrix whose rows are the electron vectors, and $n_{\beta}$ are arbitrary integers. This means that if we define \begin{equation}
\bs{\varphi}:=q^{-1\,T}\sigma \bs{\tf}= \Lambda \bs{\tf}\label{basis-transformation},
\end{equation}
then the quasiparticle spectrum is $e^{i\bsn\cdot \bs{\varphi}}$ with $\bsn$ arbitrary integer vectors.

Let us recapitulate the different sets of bosonic fields that have appeared at different stages of the calculation. Originally, the set of bosonic fields are $\{\phi_1,\,\phi_2,\,\phi_3,\,\phi_4\}$ [see Fig.~1(b) in the main paper]. We then transform to the eigenmodes $\{\tf_1,\,\tf_2,\,\tf_3,\,\tf_4\}$ whose properties are renormalized. In the low temperature regime [the action is given in Eq.~(\ref{low-T-action-sm})], $\tf_1$ and $\tf_4$ are localized. We are left with $\tf_2$ and $\tf_3$, which are renamed $\tf_{\tc}$ and $\tf_{\tn}$, and are akin to $\phi_{\rho}$ and $\phi_{\sigma}$ in the KFP theory \cite{Kane1994sm}. We then use the $\tf_{\tc}$ and $\tf_{\tn}$ to form the ``reincarnated'' ``bare'' modes $\varphi_1$ and $\varphi_2$ (not to be confused with the original $\phi_1$ and $\phi_2$). These relations are summarized as \begin{align*}
& \{\phi_1,\,\phi_2,\,\phi_3,\,\phi_4\}\xrightarrow{\text{diagonalization}}\{\tf_1,\,\tf_2,\,\tf_3,\,\tf_4\}\xrightarrow[\text{onto unlocalized modes}]{\text{renormalization, and projection}}\{\tf_2,\,\tf_3\} \\
& \xrightarrow{\text{renaming}}\{\tf_{\tc},\,\tf_{\tn}\}\xrightarrow[\Lambda]{\text{linear transformation}}\{\varphi_1,\,\varphi_2\}.
\end{align*}
In terms of the $\varphi$'s, the charge vector is $\bs{t}=\Lambda^{-1\,T}\bs{\ttt}=q\sigma\bs{\ttt}=(1,1)^T$ by construction [see Eq.~(\ref{unit-charge-condition})], independent of the choice of the electron vectors. The $K$-matrix is given by $K=\Lambda^{-1,\,T}\sigma \Lambda^{-1}=q\sigma q^T$. For Models 1, 2 and 6, it is \[
\left(\begin{array}{cc}1 & 0 \\ 0 & -3\end{array}\right),\;\left(\begin{array}{cc}1 & 3 \\ 3 & -3\end{array}\right),\;\text{and }\left(\begin{array}{cc}-3 & 6 \\ 6 & -3\end{array}\right)\]
respectively. Now we remember that the edge theory's topological order is inherited from the bulk topological order, which should not be changed by localization transition of the reconstructed edge. Hence, we conclude that the low energy fixed point is described by Model 1 above since it has the same $K$-matrix as the unreconstructed edge. The quasiparticle spectrum is, in terms of $\tf_{\tc}$ and $\tf_{\tn}$, $\exp\{i[\frac{1}{\sqrt{6}}(3n_1-n_2)\tf_{\tc}+\frac{1}{\sqrt{2}}(n_1-n_2)\tf_{\tn}]\}$, $n_1,\,n_2\in \ZZ$.

{\it Tunneling from a Fermi liquid.} Now, among other things, we can determine the $I-V$ characteristics for tunneling from a Fermi liquid to the edge. The tunneling charge $Q=\sqrt{\frac{2}{3}}\frac{1}{\sqrt{6}}(3n_1-n_2)=n_1-\frac{n_2}{3}$ has to be an integer. The contribution to the scaling dimension from the quantum Hall edge is $\frac{1}{2}\{[\frac{1}{\sqrt{6}}(3n_1-n_2)]^2+[\frac{1}{\sqrt{2}}(n_1-n_2)]^2\}|_{n_2=3(n_1-Q)}$. The Fermi liquid side can be modelled as a $\nu=1$ quantum Hall edge \cite{Chamon1997}, so the contribution to the scaling dimension is $\half Q^2$. Minimizing the sum of the two terms (call it $g$) with respect to integer $Q$ and $n_1$, we find $\min(g)=1+\half=\frac{3}{2}$, and \begin{equation}
I\propto V^{2g-1}=V^2\quad \text{(low temperature fixed point)} \end{equation}
for $Q=1$ and $\bsn=(1,0)^T$ or $(2,3)^T$, i.e. the most relevant tunneling operators are $e^{i\varphi_1}$ and $e^{i(2\varphi_1+3\varphi_2)}$, or in terms of the eigenmodes $\exp[i(\sqrt{\frac{3}{2}}\tf_{\tc}\pm \frac{1}{\sqrt{2}}\tf_{\tn})]$.

By contrast, if one tunnels an electron from a Fermi liquid to the outermost edge channel at the intermediate fixed point, the scaling dimension is $g=\half (\sqrt{3})^2+\half=2$, so the $I-V$ characteristics is \begin{equation}
I\propto V^3\quad \text{(intermediate fixed point)}.
\end{equation}

\begin{figure}
\begin{center}
\subfigure[]{\label{fig:low_V}\includegraphics[scale=0.8]{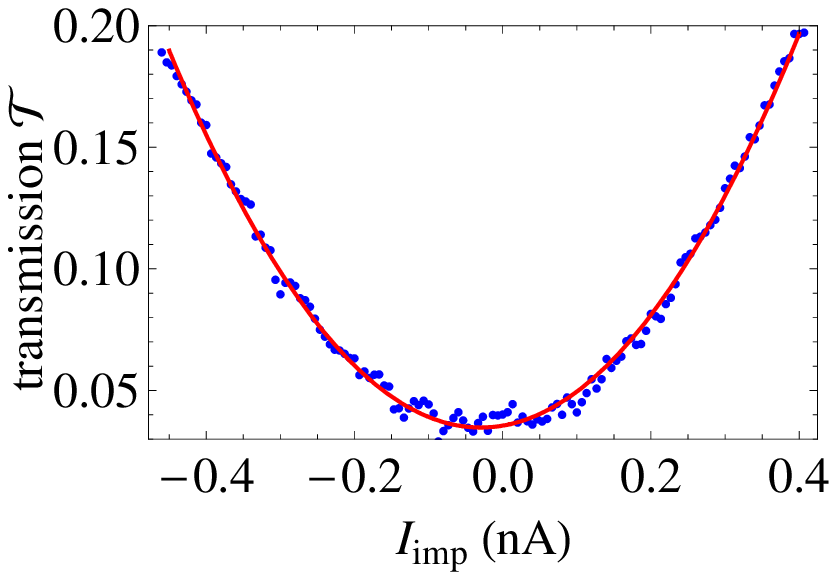}}\hspace{0.5cm}
\subfigure[]{\label{fig:high_V}\includegraphics[scale=0.8]{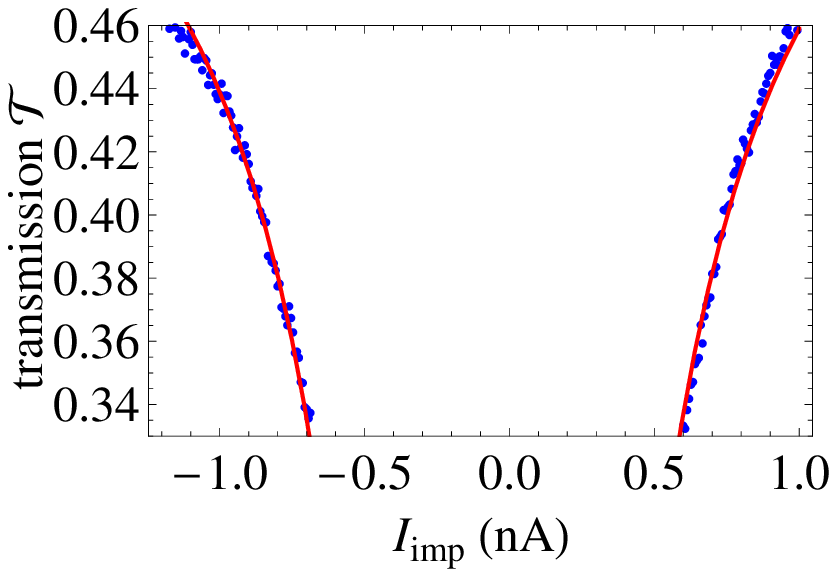}}
\end{center}
\caption{\label{fig:transmission} (Color online) Transmission $t$ at low temperature for some unspecified split-gate voltage $V_G$. The dots are original data as in Fig.~4 (upper panel) of Ref.~\cite{Bid2009sm} (we thank M. Heiblum for providing it). (a) The low impinging current regime, the solid line is a power law fit with exponent 2. We allow small horizontal and vertical shifts accounting for experimental errors in determining the origin (residual/stray currents), so the fitting formula is $\mathcal{T}=a(I_{\text{imp}}[\text{nA}]-b)^2+c$. The best fit is $a=0.877$, $b=-0.0297$, $c=0.0348$. (b) The relatively high impinging current regime, the solid line is a constant minus a power law with the exponent $-4/3$. Again we allow small shifts, so the fitting formula is $\mathcal{T}=1/2-a' |I_{\text{imp}}[\text{nA}]-b'|^{-4/3}+c'$. The best fit is $a'=0.146$, $b'=-0.0512$, $c'=0.0955$.
}
\end{figure}

{\it Transmission at a QPC.} Fig.~4 (upper panel) of Ref.~\cite{Bid2009sm} demonstrates that for some split-gate voltage $V_G$ the transmission starts as a power law at low impinging current (or low voltage) and then saturates at $1/2$ for high voltage. The low voltage data is well described by the KFP fixed point, which predicts that the transmission scales at $V^2$ [\ref{fig:low_V}]. On the other hand, the KFP theory would predict that at high voltage the transmission saturates at unity, with a correction that goes like $V^{-2/3}$, in clear contradiction to the data. Our theory, on the other hand, claims that the high voltage data is described by a different fixed point that support an outer edge state of charge $1/3$. If the 3 inner edge states are fully reflected, and the backscattering is only due to the outer edge state, the transmission approaches $1/2$, with a correction that behaves like $V^{-4/3}$, in excellent agreement with the data [\ref{fig:high_V}]. 
\end{subsection}
\end{section}

\end{widetext}

\begin{thebibliography}{28}
\bibitem [{foo()}]{footnote3}%
  \BibitemOpen
  \href@noop {} {}\bibinfo {note} {The power law behaviour is for the weak
  tunneling limit, which for the intermediate temperature regime and $eV\gg
  k_{\text{B}}T$ requires very small bare tunneling amplitude.}\BibitemShut
  {Stop}%
\end{thebibliography}
\begin{thebibliography}{11}
\bibitem [{foo({\natexlab{a}})}]{footnote1}%
\bibinfo{note}{This means that $[O_{\bsn}(x),O_{\bsm}(x')]$ vanishes for $n=m$,
  but not necessarily for $n\neq m$.}
\bibitem [{foo({\natexlab{b}})}]{footnote2}%
\bibinfo{note}{We known how to simultaneously diagonalize a real symmetric
  matrix, namely $K'$, and a real symmetric positive definite matrix, namely
  $V$.}
\end{thebibliography}
\end{document}